# An Energy Efficient Neighbour Node Discovery Method for Wireless Sensor Networks

V. Karthikeyan,
Department of ECE
SVS College of Engineering
Coimbatore, India
Karthick77keyan@gmail.com

A. Vinod[*]
Department of ECE
SVS College of Engineering
Coimbatore, India
vinodnash@gmail.com

P. Jeyakumar
Department of ECE
Karpagam University
Coimbatore, India
jeyak522@gmail.com

*Abstract*— The discovery of neighbouring nodes in multihop wireless networks has become a key challenge. Due to tribulations in communication, synchronization loss between nodes, disparity in transmission power etc, the connectivity of nodes will always experience disruptions. On the other hand, the energy utilization by the nodes also became critical. In this paper, we propose a new method for neighbour discovery in wireless sensor networks (WSNs) which pays an eminent consideration for energy utilization and QoS parameters like latency, throughput, error rate etc. In the proposed method, the network routing is enhanced using AOMDV protocol which can accurately discover the neighbour nodes and power management with HMAC protocol which reduces the energy utilization significantly. A complete analysis is being performed to estimate how the QoS metrics varies in various scenarios of power consumption in wireless networks.

## I. INTRODUCTION

Recent researches in wireless technology focus towards making the devices more sophisticated and portable. The applications of wireless networks became momentous in many areas like military scrutiny, oceanographic studies, mine discovery etc. In the network, the sensors will be deployed on a large area and the data collected by the sensor has to be transmitted from the source to the sink with maximum accuracy and least power consumption. Since recharging of power sources of the nodes is intricate, there should be a proficient energy cutback mechanism. On the other hand, for successful communication of sensor nodes in multihop sensor networks the discovery of neighbour nodes is indispensable. The nodes in the network acts as routers, which transmits data packets from one neighbouring node to another. Most of the sensor networks consist of both static and mobile nodes. Many approaches have been proposed recently for neighbour node discovery. But they are not capable to muddle through the tribulations like frequent addition of new nodes, loss of wireless connectivity, augment in transmission power etc.The most essential prerequisite of a wireless network is efficient routing of information from a source to the desired destination. For this each node should maintain the Neighbourhood information locally. Such information proviso is maintained even in mobile networks also for tracking and other docking applications. As the number of pre-positioned wireless devices become greater than before, the distribution of communication channel became a major concern. In particular for intense networks, the collision of data packets lead to the drop in throughput so that there will not exist any significant network harmonization. In this scenario, the precise assessment of neighbour nodes becomes much pertinent. This paper examines, the most imperative hinder of energy consumption in wireless networks namely neighbour node discovery and formulate an efficient protocol with which the nodes can estimate the exact information of their neighbours even if they hold on to low power mode.

On the other hand, the overall power consumption is being abridged with incorporating much more enhanced power management schemes. With this we can amplify the number of neighbour discoveries per unit time and also allow locating the mobile nodes in the network. Earlier approaches focus on static interconnected networks and experiences bootstrap problem when the nodes desires to begin at the same time. Traditional reactive protocols are not sufficient for managing the error prone behaviour of multichip dynamic networks. They make use of single route strategy for a pair of source and destination node. Due to the dynamic configuration of network the possibility of recurrent node failures are high. This makes the predefined routes unsound. Finding alternative paths in case of failures will augment the delay in packet delivery. The problem of neighbour node discovery in wireless networks can be managed significantly by the method we proposed. The overall power utilization has reduced by introducing momentous variations in wake/sleep modes of every nodes. The nodes in the networks are associated with a default routing table which consists of all route information in the network. On demand routing protocols are found to be more beneficial for dynamic networks. In the proposed method, Adhoc On demand Multipath Distance Vector Routing (AOMDV) is used to enhance default dynamic routing schemes. Whenever there are multiple paths accessible, the route to be pursued preferentiality will depend up on the administrative distance value (AD) assigned to each protocol or the route and selected route should of least power consumption. The rest of the paper is organized as follows. In the section II we review and investigate some of the recent works on neighbour node discovery. Section III analyses the neighbour discovery problem in various scenarios of sensor networks. Section IV gives the detailed explanation of our proposed method and finally in section V we conclude the paper. A typical neighbour node discovery scenario is shown in Figure 1

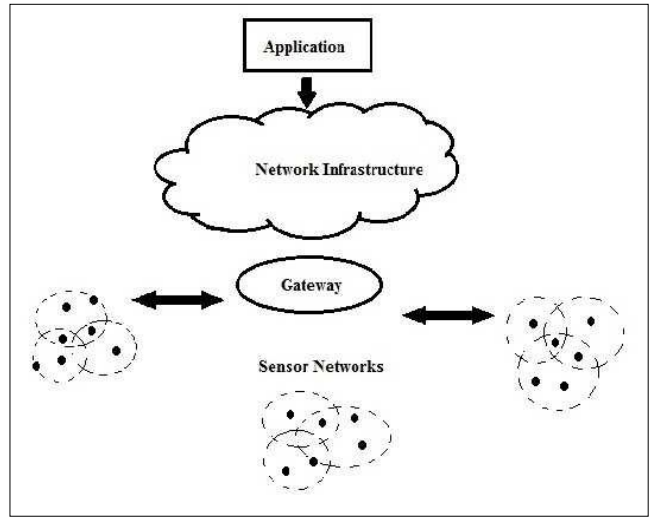

Fig. 1 A distinctive scenario of node discovery in a sensor network

## II. RELATED WORK

Researches for competent neighbour node discovery protocols have become lively over many years. Many algorithmic and realistic methods have been proposed recently. Most of them were focusing towards the accuracy of node detection. But for multihop sensor networks, the power utilization problem is more critical. In centralized networks, an extraordinary node will be there, called access point which coordinates the sharing of the medium. All the messages will be transmitted through this node only. Thus the neighbour node discovery is the uncovering of a new node in the network. Conventional method of sending HELLO messages to all nodes is not apposite for networks with restricted power supply or networks for which recharging is a big deal [1]. As the no of nodes in the networks increases the broadcast rate of HELLO messages will rise. Therefore the overall energy consumption cannot be manageable. Dutta *et al* [2] proposed a sensible solution to the neighbour node discovery problem which is based on the radio wake time scheduling of nodes. It initializes the wake times by multiples of prime numbers in such a way that the sum of their reciprocals is exactly equal to the radio duty cycle of a specific application. The node will wake up for one counter period and increments a local counter if the local counter is divisible by any of the prime numbers. The algorithm was found to be efficient for communication between a single source and destination. When more than one node is desired to communicate with a single destination, the algorithm fails. Figure 2 shows the failure of discovery algorithm. Galluzzi *et al* [3] surveyed many protocols for low-power operation in sensor networks and formulated a new with randomized schedules, deterministic schedules of wake/sleep and sophisticated methods to ensure discovery of neighbour nodes. Hamida *et al* [4] analyzed the problem of neighbour node discovery for adhoc networks. Inspired by aloha, they Proposed a HELLO protocol. In this protocol all the nodes will be either in listening mode or in talking mode. Transmission of HELLO messages is initiated by a node randomly.

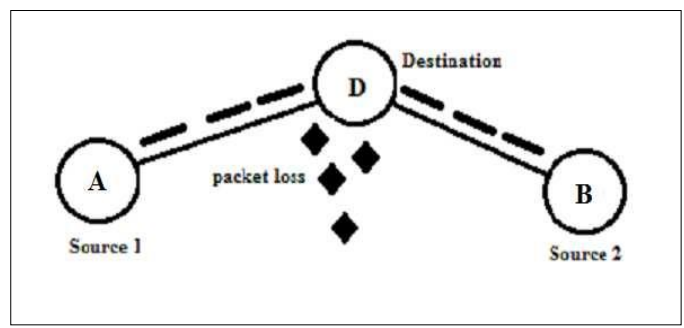

Fig. 2 Failure of node discovery algorithm

It is considered that a node is successfully discovered, if the message does not get collided with other HELLO messages. The protocol is significantly superior in estimating the transmission frequency of HELLO messages and discovery process duration. Giruka et al [5] studied the impact of HELLO protocol strategies in wireless networks. So as to reduce the congestion in network and to increase the overall performance, three HELLO protocols were proposed. In [6], Chakeres *et al* proposed to exploit of advantages of adhoc on demand routing protocol and examined the efficacy of HELLO messages for monitoring the node connectivity. The accuracy of these messages is evaluated and calculated various factors that influence the utility of them. The root of using HELLO messages is to resolve the connectivity rate from the supposition that reception of HELLO messages will point out a feasible communication channel within the source of HELLO. The method works sound on wired networks, ere packet losses are prone and probabilities of link connectivity failures are high.

The most significant problem of multihop wireless networks is successful and accurate transmission of data packets to the dynamic nodes where there is lack of a centralized control and topology is not pre-defined. On the other hand, there will be routing protocol message overhead which consumes more channel bandwidth and power of node batteries for processing and communication. Hence it will be Beneficial to use on-demand or reactive protocols in which the communication paths are established when needed. To determine a path between a source and destination, it initiates the path discovery mechanisms. The routes thus obtained are then maintained, thus putting down the control overhead and minimizing the load of the network. In our proposed method, the routing protocol message overhead is efficiently managed by implementing Adhoc on- demand multipath distance vector routing [7] which is the most suitable protocol for multihop sensor networks. On the other hand the high energy utilization of nodes is reduced significantly by Hybrid MAC protocol [8, 9].

### III. PROBLEM STATEMENT

A typical wireless sensor network consists of both static and mobile nodes. The location of mobile nodes will change continually by time. As a result, the topology of the network cannot be pre-defined. Hence the connectivity between the nodes will get disturbed and nodes moves outside the range of communication. Neighbour node discovery problem is nothing but detecting the mobile nodes within one node's communication range. The location information of nodes over time has to be updated accordingly. Also multiple numbers of nodes should not be allowed to access the same destination at the same time there by avoiding packet loss and errors. One of the eminent problems of MANETs termed as hidden node problem, which is being solved using RTS/CTS messages, cannot be adopted for solving the same in sensor networks. Because, stations are in sensor networks are either primary or secondary. In sensor networks incorporation of combined stations is very costly.

Networks in which the nodes are not synchronized with each other will be detached frequently. Due to dynamic behaviour of nodes the location information also will be changed. The nodes can communicate efficiently only if there is a stable wireless connectivity between them. In most of the networks the power source for the nodes are batteries. Due to many practical limitations, the recharging of these batteries is very difficult. Hence the energy consumed by the nodes should be condensed as much as possible. The power used for communication and processing of messages has to be minimized to the maximum. On the other hand a large amount of energy is exhausted while nodes are idle. During the node listening period, the battery power will be wasted unhelpfully.

### IV. PROPOSED DISCOVERY METHOD

In this section, we present our neighbour node discovery method which is efficient in tracking the nodes with in a node's communication range. The study performance analysis of common node discovery protocols has been performed in due course. During the analysis we considered the network as unit graph, in which any pair of nodes comes under transmission range are neighbour nodes. The reduction of overall power consumption is managed by introducing significant variations in wake/sleep times of the nodes. Since the network is dynamic in nature, the routing protocol will be preferred on the basis of administrative distance value allotted to each path in the network. This will appreciably reduce the end to end delay and routing protocol overheads. The explanation of our proposed approach can be divided in to three subsections as mentioned below.

*A. Sensor Network with Dynamic Source Routing*

A synchronized sensor network can initiate efficient neighbour node discovery using dynamic source routing protocol. In DSR, all the routing information is maintained and updated continually at the dynamic nodes and it is independent of routing tables of intermediate nodes. The route establishment and maintenance are performed by separate mechanisms. The neighbour discovery in the network is done as follows. First of all the source node will flood a number of Route Request packets to all other nodes in the network. When a node receives it, a Route Reply packet will be transmitted back to the source. The sequence number of each packet avoids loop formation and multiple broadcast of same Route Request. The advantage of using DSR protocol is that, there is no need of HELLO messages. This protocol can be adopted when the disruptions to the node connectivity is fewer. When the connectivity strength is less, DSR will reconfigure the routes and becomes stable after a while. But if any node in the route fails, the data transmission will be unsuccessful.

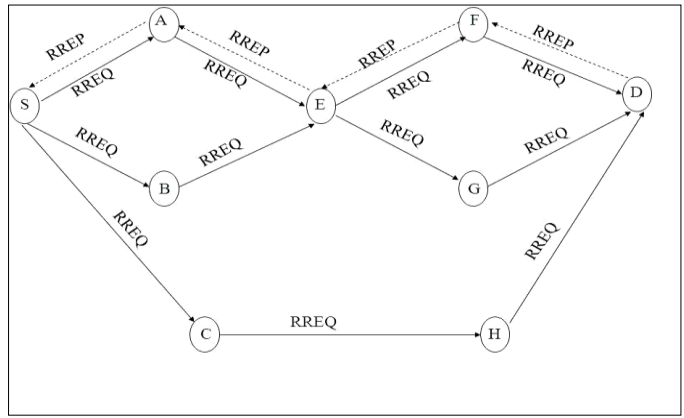

Fig. 3 Message transmissions in AOMDV protocol

*B. Sensor Network with Adhoc On-demand Multipath Distance Vector Routing*

The disabilities of DSR protocol are efficiently managed by Adhoc On-demand Multipath Distance Vector Routing protocol. This protocol is an enhancement of traditional on- demand protocol named as AODV protocol. AOMDV is sound in formulating multiple paths which are loop free and disjoint. The routing entries for each destination are maintained in such a way that it consists of a list of next hops along with corresponding hops. Same sequence number is allotted to next hops. This avoids loop formation. The destinations will send out route advertisements. This is done based on the advertised hop count which is maintained. Advertised hop count is the maximum hop count for all the paths. Figure 3 shows how the source reaches the destination the discovery of neighbour nodes is initiated by Route Discovery mechanism of AOMDV protocol. During this session, the source node will flood of Route Request (RREQ) messages throughout the network.

The node which is intermediate to source node will receive the RREQ and sets up a reverse path. This is done by making the previous hop of the request message as the next hop on the reverse path. Multiple numbers of RREQ messages reached at the same node will be discarded. This will significantly reduce the network load and delay in the network. As the RREQ message reaches the destination, a Route Reply (RREP) message is generated and sends back to the source node through the same path followed by the RREQ message. The next session is the Route Maintenance. Whenever the connection between any two nodes is damaged or lost, a Route Error (RERR) message is generated and sends to all sources through precursor routes which are maintained discretely. Pre-defined routes are erased by RERR messages accordingly. When a source receives a RERR message, it initiates Route Discovery session again.

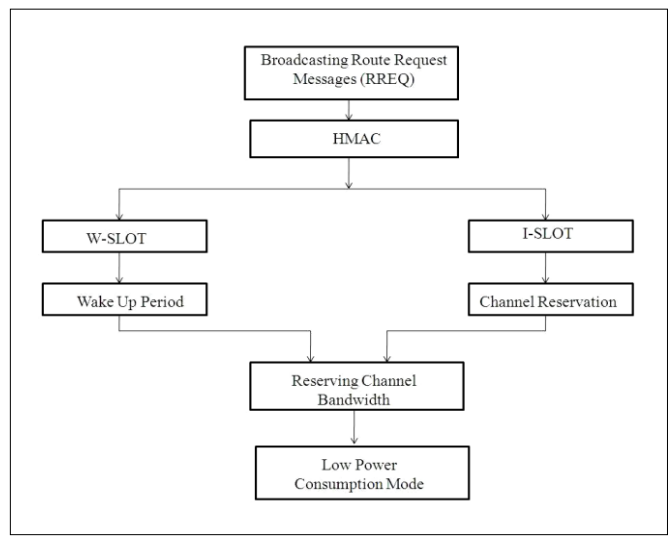

Fig. 4 Implementation of HMAC protocol

*C. Energy management using Hybrid MAC protocol*

In Hybrid MAC protocol, the time is alienated in to two slots namely Wakeup Slot (W-SLOT) and Information Slot (I- SLOT). W-SLOT is very shot in time, while I-SLOT is subdivided in to a number of slots. Every node in the network is assigned a unique W-SLOT so that it can listen to other wakeup messages. Figure 4 shows the implementation of HMAC protocol. The nodes will be in sleep state during all other wakeup slots. Whenever a node desires to transmit data, it will arbitrarily pick an I-SLOT and indicates the receiver node(s) with corresponding slot number in the receivers W- SLOT through a wakeup message. Thus only that receiver wakes in the corresponding I-SLOT for data reception. During this time, all the other nodes will be in sleep mode. Hence the energy utilization by the nodes in the network will reduce drastically. HMAC protocol supports one hop broadcasting also. Whenever a node desires to broadcast information, it will send a wakeup message including the address of broadcasting and an information slot in each wakeup slot. Once the wakeup message is received, all the neighbour nodes will wake up in the same I-SLOT and receives the broadcast message.

## V. CONCLUSION

We analyzed the neighbour node discovery problems in static and dynamic multihop Sensor Networks. Conventional node discovery techniques were found to be inadequate and they give less significance to the QoS parameters. So as to overcome these issues we proposed a new method for node discovery in which AOMDV protocol is implemented for node discovery and route establishment. The comparative study with other routing protocols, we came to the inference that, our method can produce high discovery rate. On the other hand, for reducing the overall power utilization Hybrid MAC protocol is implemented. Results show significant reduction in power consumption with HMAC. Hence the end to end delay and routing over head and energy consumption are minimized and the throughput is increased remarkably.